\documentclass{article}

\usepackage{arxiv}

\usepackage[utf8]{inputenc} % allow utf-8 input
\usepackage[T1]{fontenc}    % use 8-bit T1 fonts
\usepackage{hyperref}       % hyperlinks
\usepackage{url}            % simple URL typesetting
\usepackage{booktabs}       % professional-quality tables
\usepackage{amsfonts}       % blackboard math symbols
\usepackage{nicefrac}       % compact symbols for 1/2, etc.
\usepackage{graphicx}
\usepackage{doi}
\usepackage{linguex}
\usepackage{colortbl}
\usepackage{tabulary}
\usepackage{xstring}
\usepackage{xcolor}
\usepackage{float}

\title{Predicting the winner of the US 2024 elections \\using trust analytics  }

\author{ 
\href{https://orcid.org/0000-0001-9674-9902}{\includegraphics[scale=0.06]{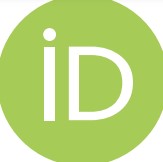}\hspace{1mm}Katarzyna ~Budzynska} \\
	Laboratory of The New Ethos\\
	Warsaw University of Technology, Poland \\
	\texttt{katarzyna.budzynska@pw.edu.pl} \\
 	\And
\href{https://orcid.org/0009-0006-6012-4787}{\includegraphics[scale=0.06]{orcid.jpg}\hspace{1mm}Ewelina ~Gajewska} \\
	Laboratory of The New Ethos\\
	Warsaw University of Technology, Poland \\
	\texttt{ewelina.gajewska.dokt@pw.edu.pl} }

\renewcommand{\headeright}{}
\renewcommand{\undertitle}{}
\renewcommand{\shorttitle}{Predicting the winner of the US 2024 elections using Trust Analytics}

\hypersetup{
pdftitle={Predicting the winner of the US 2024 elections using trust analytics},
pdfsubject={ },
pdfauthor={Ewelina ~Gajewska},
pdfkeywords={social media, trust analytics, BERT, detection, elections, ethos },
}

\begin{document}
\maketitle

\begin{abstract}
A number of models and techniques has been proposed for predicting the outcomes of presidential elections. Some of them use information on the socio-economical status of a country, others focus on candidates' popularity measures in news media. We employ a computational social science approach, utilising public reactions in social media to real-life events that involve presidential candidates. Contrary to the popular approach, we do not analyse public  emotions but \textbf{ethotic references} to the character of politicians which allows us to analyse how much they are (dis-)trusted by the general public, hence the name of the tool we developed: \textbf{Trust Analytics} (TrustAn). Similarly to major news media's polls, we observe a tight race between Harris and Trump with week to week changes in the level of trust and distrust towards the two candidates. Using the ratio between the level of trust and distrust towards them and changes of this metric in time, we predict \textbf{Donald Trump} as the winner of the US 2024 elections. 
\end{abstract}

% keywords can be removed

%\section{Introduction}

\section{Methodology: Data and mining}
Material for the analysis comprises of discussions on Reddit  about political issues inside the r/politics subreddit community. Every month, from June to October 2024, we downloaded around 15-20 key discussions.  In total, we collect 87 discussions from the last 5 months preceding the election day on November 5th, 2024. 
Each Reddit post is split into sentences using regular expressions (``?'', ``!'', ``.'') and filtered for the occurrence of the candidates' names. 
This preprocessing step leaves us with 101,482 sentences for the analysis.

Such preprocessed data is fed through our ethos mining model to identify statement which support and attack the ethos of Trump and Harris. Our ethos mining model \cite{pprai} is an updated version of \cite{duthie2016mining, duthie2018deep}, curated for the specificities of language expressions in social media. It classifies sentences into ternary categories of: attack, support or no ethotic expression. Such a taxonomy of labels is built upon the Aristotelian concept of rhetoric \cite{AristotleRhet}. Support and attack of one's ethos is defined then as a favourable and unfavourable, respectively, reference to an entity, in particular praising their wisdom, virtue or goodwill (support) and questioning their credibility through, for example, association of the entity with events of a negative connotation. Our RoBERTa-based model achieves 0.758 macro-$F_1$ on a test set comprising polarised discussions on social issues from Twitter and Reddit platforms \cite{polaris}.

\section{Results: Analytics}
Trust Analytics, TrustAn, has been developed to statistically analyse the result produced by  the ethos mining technique on the Reddit data. This tool is a module of the Rhetoric Analytics platform  \cite{Budzetal2024tne} that allows for the analysis of the strategic use of language in the digital society. 

\begin{figure}[h]
    \centering
    \includegraphics[width=0.8\linewidth]{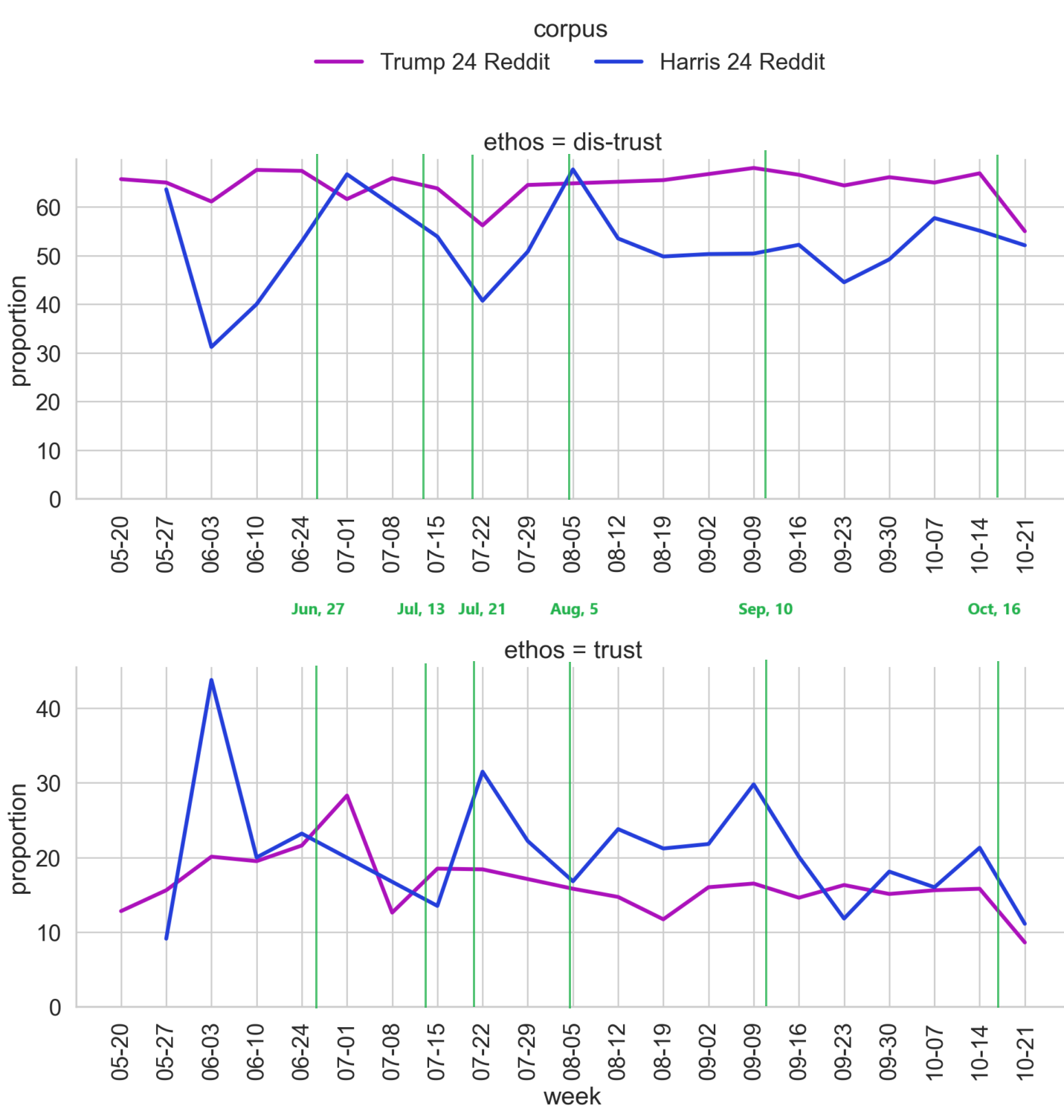}
    \caption{\textbf{Proportion}. Changes in the raw level of trust and distrust towards the candidates.}
    \label{fig:summary}
\end{figure}

As the concept of ethos is associated with trust and credibility of a speaker, henceforth we refer to these attack and support labels as \textit{distrust} and \textit{trust}, respectively. 
A basic distribution of the level of distrust and trust towards Harris and Trump can be found in Figure \ref{fig:summary}. In addition to the time scale, we mark several events of high importance in the context of elections, that is: a presidential debate between Joseph Biden and Donald Trump (June, 27th); attempted assassination of Donald Trump in Pennsylvania (July, 13th); Biden's withdrawal from the presidential race (July, 21st); official announcement of Kamala Harris as the Democratic nominee for the US 2024 presidential elections (August, 5th); presidential debate between Kamala Harris and Donald Trump (September, 10th); and Kamala Harris's interview in Fox News station (October, 16th). These are marked by vertical lines in Figure \ref{fig:summary}.
Overall, throughout the analysed period of time we observe a higher level of distrust and a lower level of trust towards Trump than Harris. %First presidential debate did not substantially changes the polls for Trump. A slight increase of trust and decrease of distrust towards him is observed after the attempted assassination on July, 13th. Announcement of Biden's withdrawal from the presidential race increases the level of trust towards Harris in the following week. 

\begin{figure}[h]
    \centering
    \includegraphics[width=0.85\linewidth]{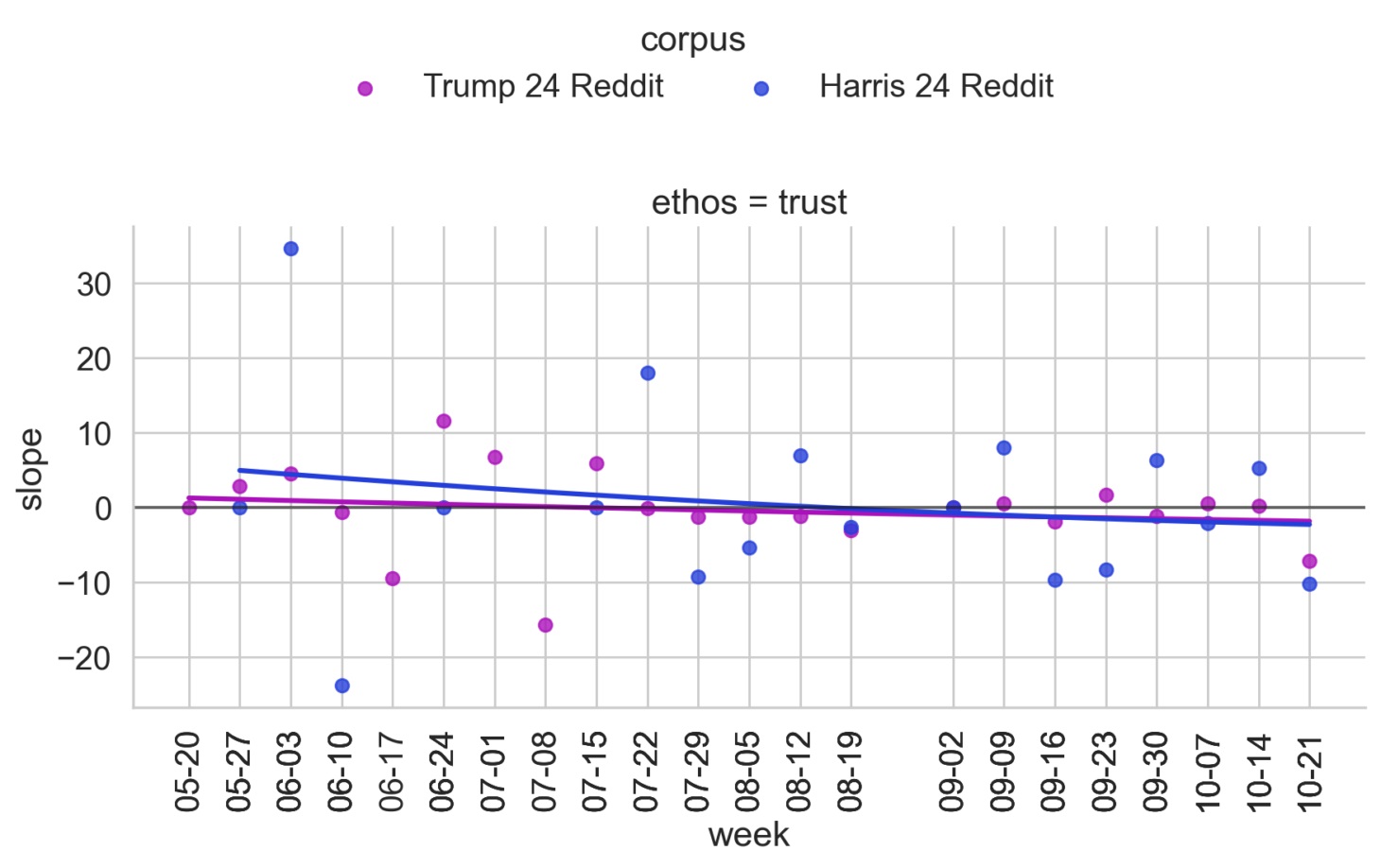}
    \caption{\textbf{Slope}. Week-to-week slope in the level of trust towards the candidates.} % Trump vs. Harris 2024 polls: 
    \label{fig:slope}
\end{figure}

Based on results of this automatic annotation, we calculate two metrics for computing the distribution of support for Harris and Trump in the last 5 months of election campaigns. 
Lineplot in Figure \ref{fig:slope} 
presents a weekly changes of trust towards the candidates using the calculation of a slope: $s_1 = (y_{t1} - y_{t0}) / (x_{t1} - x_{t0})$ (where $x_{t1}$ and $x_{t0}$ are the time points of the current week  and one week earlier; and $y_{t1}$ and $y_{t0}$ are the values of trust in these weeks).
Here, we observe two things. First, week-to-week levels of trust towards Trump are rather stable throughout the analysed period of time. Second, since the beginning of September levels of trust towards Harris and Trump are very tight (both with a decreasing tendency), while October shows weekly changes in the lead of the presidential race. As a result, this metric cannot reveal the winner of the US 2024 elections.

\begin{figure}[h!]
    \centering
    \includegraphics[width=0.83\linewidth]{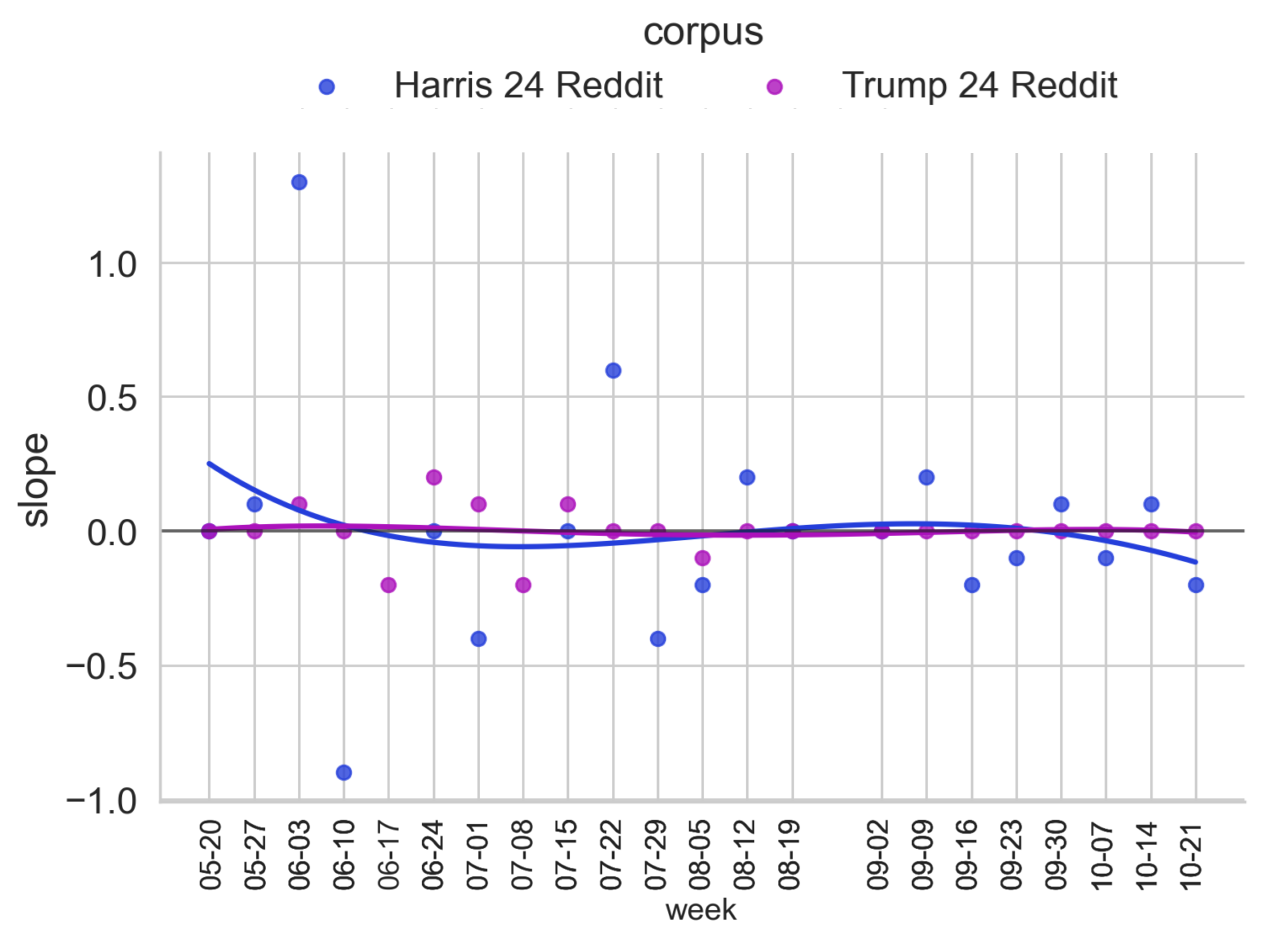}
     \caption{\textbf{Profile}. Week-to-week slope in the ratio between trust and distrust towards the candidates.}
    \label{fig:sloperatio}
\end{figure}

Thus, we created another metric of \textit{trust profiles} which  calculates ratio between the level of trust and distrust and its weekly changes using the formula provided above. Results are displayed in Figure \ref{fig:sloperatio}. 
Again, we observe a stable profile of Trump in the last two month preceding the elections. Harris's trust profile fluctuates from month to month. It initially showed an increase from August to September, giving her more credibility than Trump. Since the end of September, however, her profile becomes negative and further decreases till the end of October. 
Given a stable profile of Trump and a decreasing tendency in the trust profile of Harris in the last month of the presidential race, we predict Donald Trump as the winner of the US 2024 elections. As  Figures \ref{fig:summary}-\ref{fig:sloperatio} demonstrate: although Trump's sceptics are loud, his supporters are louder. 

% trump has a lot of haters but also a lot of supporters - Trump's opposition is loud but his supporters are louder. 
% we note a more stable public trust profile for Trump compared to Harris

%%%%%%%%%%%%%%%%%%%%%
\nocite{*}
\section{Acknowledgements}

We thank Marcin Koszowy and Maciej Uberna  who significantly 
contributed to the research and the approach reported in this paper. We would like to acknowledge that the work reported in this paper has been supported by the Polish National Science Centre, Poland (Chist-Era IV) under grant 2022/04/Y/ST6/00001.

%\vspace{3cm}

%\bibliographystyle{unsrtnat}
%\bibliography{references}  
%%% Uncomment this line and comment out the ``thebibliography'' section below to use the external .bib file (using bibtex) .


\end{document}